**Erratum: Phononic soft mode and strong electronic background behavior across the structural phase transition in the excitonic insulator Ta2NiSe5 [Phys. Rev. Research 2, 042039 (2020), arXiv:2007.01723v2]**

Min-Jae Kim[1,2], Armin Schulz[1], Tomohiro Takayama[1,3], Masahiko Isobe[1], Hidenori Takagi[1,3,4], and Stefan Kaiser[1,2]

[1] Max Planck Institute for Solid State Research, Stuttgart, Germany
[2] 4th Physics Institute, University of Stuttgart, Germany
[3] Institute for Functional Matter and Quantum Technologies, University of Stuttgart, Germany
[4] Department of Physics, University of Tokyo

After publication of [Phys. Rev. Research 2, 042039 (2020), arXiv:2007.01723v2] [1] we got notified on an instrumental artifact in the low frequency part of the Raman spectrum of $Ta_2NiSe_5$. This erroneously led to a description of a soft phonon mode and a strong electronic background. Properly taking into account a low frequency electronic signal correctly explains the observed soft mode as Fano coupled phonon where a critical softening of an excitonic collective mode takes place as described by P.A. Volkov et al. in arXiv:2007.07344 [2] and M. Ye et al. in arXiv:2102.07912 [3].

In our publication [1] we do describe the Raman spectrum of the excitonic insulator Ta2NiSe5 over a large temperature range from 10-800K. We got notified that the low frequency spectrum is suppressed by instrumental frequency filters in the Raman instrument.

At high temperatures above the phase transition spectral weight is pushed into this low frequency regime. In our analysis this erroneously led to a wrong description: We described the red background feature in Figure 3 as an independent electronic feature from the observed soft mode in blue that we attributed to a soft phonon.

Taking into account a correct low frequency spectrum as shown in Refs. [2,3] shows that the low frequency response is described by a Fano coupled phonon (e.g. see Fig. 6 in ref [3]) that does not soften. The observed softening of the peak is due to a critical softening of an excitonic collective mode.

The suppression of the low frequency background in our measurement is also likely the cause of the anomaly in laser heating. As Ref. [3] points out there is no different heating between the electronic and phononic system (see Fig. 13 in Ref. [3]).

Therefore the coupled phononic-excitonic interplay across the phase transition in Ta2NiSe5 is rather driven by the excitonic system as described in Refs. [2,3].

We thank G. Blumberg to point out the problems in our low frequency Raman response and the discussion on instrumental artifact.

[1] M.-J. Kim et al. Phys. Rev Research **2**, 042039 (2020).
[2] P.A. Volkov et al. arxiv:2007.07344 (2020).
[3] M. Ye et al. arxiv:2102.07912 (2021).

# Phononic soft mode and strong electronic background behavior across the structural phase transition in the excitonic insulator Ta2NiSe5

Min-Jae Kim[1,2], Armin Schulz[1], Tomohiro Takayama[1,3], Masahiko Isobe[1], Hidenori Takagi[1,3,4], and Stefan Kaiser[1,2]

[1] Max Planck Institute for Solid State Research, Stuttgart, Germany
[2] 4th Physics Institute, University of Stuttgart, Germany
[3] Institute for Functional Matter and Quantum Technologies, University of Stuttgart, Germany
[4] Department of Physics, University of Tokyo

mj.kim@fkf.mpg.de, s.kaiser@fkf.mpg.de

Ta2NiSe5 became one of the most investigated candidate materials for hosting an excitonic insulator ground state. Many studies describe the corresponding phase transition as a condensation of excitons breaking a continuous symmetry. This view got challenged recently pointing out the importance of the loss of two mirror symmetries at a structural phase transition that occurs together with the semiconductor – excitonic insulator transition. For such a scenario an unstable optical zone-center phonon at low energy is proposed to drive the transition. Here we report on the experimental observation of such a soft mode behavior using Raman spectroscopy. In addition we find a novel spectral feature, likely of electronic or joint electronic and phononic origin, that is clearly distinct from the lattice dynamics and that becomes dominant at Tc. This suggests a picture of joint structural and electronic order driving the phase transition.

*Introduction.* Recently the possibility of realizing the elusive state of an excitonic insulator (EI) in the zero-gap semiconductor Ta2NiSe5 (TNSe) has stimulated a tremendous body of experimental and theoretical work. An EI consists of condensed excitons forming a new phase of matter whose proposed macroscopic quantum states may resemble many properties of superconductors and superfluids [1-8]. Besides this interesting test bed for many-body physics, the ultrafast response in TNSe under laser-excitation is a potential platform for novel optoelectronic applications [9,10]. An EI transition by spontaneous condensation of excitions is expected in narrow bandwidth semimetals or small bandgap semiconductors if the exciton binding energy exceeds the bandwidth or the bandgap of the system. EIs have been realized in specifically designed electron-hole bilayer systems [11-15] or by creating high exciton densities under strong photoexcitation [16-18]. However, identifying fingerprints of EI phases in bulk materials remain an experimental challenge. Prominent candidates like 1T-TiSe2 show an indirect band gap and therefore the potential EI state appears together with a charge density wave (CDW) that seems to drive the dynamics [19,20]. In contrast, TNSe is a direct band gap semiconductor that has been proposed to show an EI transition without charge, or other finite momentum order [21,22]. A semiconductor-insulator transition appears at Tc=328K that can be tuned via chemical or physical pressure [23,24]. Evidence for excitonic condensation comes from a characteristic band flattening seen in angle-resolved photoemission [25-27] and the opening of a gap in electronic [9,28] and optical [29-31] spectra. Further fingerprints stem from non-equilibrium responses of the ground state that are compatible with the melting and relaxation dynamics of a condensate [32-36] and the excitation of possible collective modes of a coherent ground state [37-41]. Nevertheless, the origin of the gap is still under debate due to a structural phase transition from a high-temperature orthorhombic to a low-temperature monoclinic phase that happens simultaneously at Tc despite the absence of CDW [22,42]. TNSe crystallizes as a quasi-one-dimensional structure of Ta-Ni-Ta chains. In the low-temperature phase two mirror symmetries of the high-temperature phase are broken by a shear motion of the Ta against the Ni chains [Fig. 1, inset]. Using inelastic x-ray measurements Nakano et al. [42] show that this displacement can be decomposed into B2g phonons of the high-temperature phase. While they find a significant frequency jump of the transverse acoustic mode describing the Se atom motion when approaching the phase transition from the high temperature side they cannot evidence a softening in the optical phonon mode in the temperature range they accessed. The latter mode describes the Ta atom shear motion that happens at the phase transition. But a strong linewidth broadening of this mode indicates strong electron phonon coupling. A strong lattice interaction is also evidenced by strong Fano resonances [30] and polaronic bands [31] in optics or the lattice coupled dynamics of the non-equilibrium responses [32,33,36-41]. The proposed scenario for the EI in TNSe is that the excitonic condensate couples to the lattice phonons and as such drives a joint excitonic and structural instability [24,29,43]. However, this picture of exciton condensation driven by pure electronic Coulomb interactions of the electron hole pairs got recently challenged. Exciton condensation would require breaking a continuous symmetry and a complex order parameter to describe the low-temperature phase of TNSe. In particular Mazza et al. [44] and Watson et al. [45] point out the loss of two mirror symmetries in the low temperature phase are linked to a discrete symmetry break. They propose possible excitonic (linearly coupled to lattice modes) and structural instabilities that are in agreement with the experimental observations of Nakano et al. [42]. Both instabilities lead to a hybridization of the Ta and Ni bands in the center of the Brillouin zone in contrast to the pure bands in a condensation scenario. A theoretical density functional theory study by Subedi [46] has recently explored the corresponding zone center optical phonon branches relevant for this case and suggests that the optical B2g modes could drive the phase transition. According to the calculations, if the B2g phonon mode softens as the Tc is approached from above, the dynamical instability is due to an unstable optical phonon mode. No softening should be observed if the instability is electronic or due to an unstable elastic mode corresponding to a uniform shear distortion of the lattice. However clear experimental fingerprints for the dynamics of such instabilities are missing.

This motivates our Raman studies on TNSe presented here to explore the crucial at low frequency behavior. We clearly identify a B2g soft mode above Tc and characterize a broad electronic or coupled electronic-phononic feature next to the soft mode that dominates across the phase transition.

*Experiment.* Single crystals of Ta2NiSe5 were grown by chemical vapor transport reaction described in [23]. Raman measurements are performed within specific excitation geometries and parameters as detailed in the supplemental material.

*Results* - Figure 1 shows the Raman spectra of TNSe in Y(X-)$\bar{Y}$ geometry in the temperature range from 10 to 800 K. That goes clearly beyond the 400 K covered in the inelastic x-ray measurements [42]. The blue spectra are taken in the monoclinic phase below Tc and belong in the given experimental geometry to the Ag channel. Above Tc spectra are shown in red and are taken in the orthorhombic phase so that in this geometry the experiment probes the Ag and B2g channel. Spectra below Tc show clear peaks at 101, 124, 136, 149, 180, 195cm-1 identified as Ag symmetries already in earlier studies [9, 47]. On increasing temperature the high frequency modes do not show strong changes up to Tc except the modes at 124 and 136cm-1 shifting to lower energy and broaden. At the same time they decrease in amplitude. Above Tc the 124 cm-1 mode disappears while the 136 cm-1 mode shows up to the highest temperature. Increasing the temperature further shows a slight redshift and broadening of the mode. The 101 cm-1 mode above Tc shows a similar trend. This behavior is known from the previous studies [47]. Here we will focus in particular on the so far unexplored low frequency response with prominent peaks at 36cm-1 and 71cm-1 below Tc. The 36 cm-1 Ag mode is known as a peculiar mode that under strong impulsive excitation shows a coupling to the excitonic system and resembles amplitude mode behavior [37]. In agreement with reports in the non-equilibrium study, in the equilibrium measurements this mode shows basically no change of width or shift in frequency as function of temperature. In amplitude the equilibrium mode increases as function of increasing temperature in stark contrast to the non-linear excitation.

*Ag and B2g low frequency spectra*. However the 71 cm-1 mode and the spectral range right below will become the dominant key features of the transition. Below Tc, in the monoclinic phase, shown as the blue spectra in Fig. 1 the mode barely changes on increasing temperature except for a small redshift on approaching Tc. Hardly noticeable but already present below Tc is a small broad background of spectral weight in the frequency range from 42-71 cm-1. Its weight increases towards Tc. Above Tc in the orthorhombic phase, shown in the red spectra, very prominent changes happen. Across the phase transition the mode at 71cm-1 is strongly suppressed and the broad spectral feature becomes clearly visible next to this mode. On further temperature increase in the orthorhombic phase, the broad spectral feature is most prominent between 330 and 400 K before an additional peak appears on top of the broad background and becomes sharper and dominant on increasing temperature.

To understand and characterize this broad feature as well as the appearing modes we map each symmetry channel individually (Supplemental material S2). Both key features, the 71 cm-1 Ag mode and the broad new feature appear only in the XZ channel. We plot the temperature dependent spectra of this channel in Fig. 2. Below Tc (green/yellow spectra), probing the Ag channel we find the sharp phonon modes at 71 cm-1 and the high frequency modes at 101, 124, 136, 149, 180, 195 cm-1. On increasing temperature the Ag modes in the XZ configuration become suppressed and disappear at Tc. This is in full agreement since XZ probes the B2g channel above Tc. The Ag modes remain in the XX configuration at 400K that probes the Ag channel above Tc (black spectrum). In the B2g channel of the XZ geometry above Tc (orange/red spectra) new small modes appear at 96 and 150cm-1. Most prominent is the appearance of the broad background feature at temperatures up to 400 K and the additional new peak around 60 cm-1 on further temperature increase.

*Characterization of the modes*. In the following we present the fit of the temperature dependence of the spectra. All phonon modes are fit using single Gaussian functions. However the anomalous broad spectral feature is difficult to capture. It can be fit using multiple Gaussians (where the number of fit parameters becomes large) or as we have done here using a super Gaussian function. Figure 3 shows the significant low frequency Raman spectra in the XZ geometry and the extracted fit parameters for three features: (1) As key feature at highest temperatures we identify a strong sharp B2g phonon mode at 60cm-1 (blue). On decreasing temperature this mode clearly softens to 52 cm-1 and significantly broadens approaching Tc. Its intensity decreases and disappears at Tc. This soft mode we classify as the B2g zone-center optical phonon mode predicted by Subedi [46]. The inelastic x-ray study [42] did not find this unstable B2g mode in the range of 328-400K since it already softened and broadened significantly. Also within our data we cannot resolve the mode below 400 K. (2) However, at 400 K we see the onset of a new mode that evolves into the 71 cm-1 (~2 THz) Ag mode in the monoclinic phase below Tc (green) where the mode suddenly sharpens at the phase transition. In line with the interpretation of

Subedi this new mode in the monoclinic phase corresponds to the amplitude modulation of the order parameter deriving from the unstable B2g mode of the orthorhombic phase [46]. The most surprising feature in TNSe is (3) the anomalous broad spectral component that dominates the weight in the temperature range around the phase transition (red). At very high temperatures it clearly exists with a strong finite weight next to the soft mode. When cooling below 450 K its weight significantly increases and it becomes the dominant feature approaching Tc. Then its intensity significantly drops at Tc but remaining comparable to other modes even below Tc, as seen e.g. at the 300 K data. Only when cooling further down the relative weight decreases further but still finite down to lowest temperatures. As mentioned this feature is fit using a super-Gaussian contribution so that it is difficult to assign. Possibilities are fluctuations that are small in the low-temperature ordered phase that significantly increase and peak at Tc and remain very prominent even for temperatures far above Tc. As the feature also could be fit with multiple peaks it also could represent other yet unidentified phonon modes that in the high temperature phase are suppressed from the strong soft mode behavior. To gain additional information on the background feature we have performed the Raman measurement with different excitation conditions. Figure 4(a) shows that changing the excitation wavelength to 532nm reproduce the results of Figure 2 at 632nm. However, significant changes appear on heating the system with high laser power in the Raman experiment rather than using a thermal heater. The laser heating response in Figure 4(b) reveals that the broad spectral feature is highly susceptible to the heating via laser fluence while the phononic system is not. This observation indicates an electronic or joint electronic-phononic origin of the feature that links to the proposed excitonic fluctuations. Important to note is that the soft mode behavior of the B2g soft mode does not appear on laser heating. It is either fully buried in the electronic background or is suppressed as discussed in the supplemental material (S3).

To contrast the soft mode instability and the strong electronic background feature likely jointly driving the phase transition and the appearance of a new mode in the monoclinic low-temperature phase of TNSe we show the phonon mode behavior in the sister compound Ta2NiS5; a semiconductor that does not show an excitonic nor a structural phase transition [23]. Its equivalent 2 THz mode (Figure 3, grey) shows a roughly monotonic increase in intensity and only a slight red-shift on cooling across the measured temperature range. Also the mode stays always very sharp in the whole temperature range.

*Conclusion.* We have performed a detailed study of the low frequency lattice dynamics across the orthorhombic to monoclinic structural phase transition in Ta2NiSe5. We can clearly identify a B2g optical zone-center phonon soft mode in the orthorhombic phase at temperatures above Tc. This observation underlines a proposed mechanism by Subedi [46] that this instability could act as a prime cause for the structural phase transition; leading to the appearance of a new Ag mode in the monoclinic phase below Tc, which we also clearly observe. However, this change between the different modes in their specific symmetry channels seems not to happen abruptly at Tc but smeared out around the structural phase transition. Interestingly, right in that regime a broad new feature in the low frequency regime dominates that traces fluctuating electronic or fluctuating joint electronic and structural order that drive the transition. Recently two similar Raman studies appeared as preprints [48,49] that do not report any soft mode behavior but claim a pure excitonic driven phase transition based on a low frequency electronic background. Our study does not show such a significant electronic background except the broad electronic feature discussed in Figure 3. Our electronic feature also remains distinct if the data is Bose-factor corrected: As discussed in the supplemental material (S4) reasons for the different response reported in [48,49] are possible polarization leaks from other crystallographic directions and most important the laser heating of the electronic system. As we have shown high laser power induces a dominant electronic heating and therefore stronger electronic background while the B2g soft mode behavior is not observed. This heating of the electronic system also explains that high power pump probe experiments do not trigger the structural phase transition [32]. Here it explains why the two other Raman studies do not find the phononic soft mode as we report here and only see a dominant electronic response (see also supplemental discussion S3 and S4). However, the studies [48,49] identify excitonic fluctuations as nature of their electronic background. That substantiates our identification of the broad background feature being of dominant electronic origin. Excitonic fluctuations that we can link to the feature are strongest at Tc as in Ref. [49]. However in our study we observe that they not solely drive an excitonic phase transition but smear out a joint structural and electronic transition.

From this we conclude that in Ta2NiSe5 the phase transition is neither purely phononic [45,50] nor purely or dominant electronic [48,49,51] as from contradicting conclusions in various studies. The present study rather points to an interplay of both electronic and structural order. A possible scenario could be self localized exciton in polaron complexes as suggested in [30,31]. These may act as natural cavities for the excitonic system. This results in the observed high susceptibility of the electronic system to laser fluence by breaking the excitons and the only weakly affected lattice dynamics of the remaining polaronic distortion. This may also explain some of the contradicting conclusions drawn in the literature on the nature of the excitonic/structural phase transition in Ta2NiSe5 so far. As such the present study serves as a first benchmark experimentally describing a jointly driven transition and also as trigger to further theoretical and experimental studies to characterize the electronic and phononic interplay in Ta2NiSe5.

**Acknowledgements** We thank A. Subedi and A. V. Boris for insightful discussions.

**References**

[1]  N. F. Mott, Philos. Mag. **6**, 287 (1961).
[2]  Keldysh, Sov. Phys. Solid St **6**, 2219 (1965).
[3]  D. Jerome, T. M. Rice, and W. Kohn, Phys. Rev. **158**, 462 (1967).
[4]  B. I. Halperin and T. M. Rice, Rev. Mod. Phys. **40**, 755 (1968).
[5]  F. X. Bronold and H. Fehske, Phys. Rev. B **74**, 165107 (2006).
[6]  K. Seki, R. Eder, and Y. Ohta, Phys. Rev. B **84**, 245106 (2011).
[7]  B. Zenker, D. Ihle, F. X. Bronold, and H. Fehske, Phys. Rev. B **85**, 121102(R) (2012).
[8]  Z. Sun and A. J. Millis, Phys. Rev. B **102**, 041110(R) (2020).
[9]  S. Y. Kim, Y. Kim, C.-J. Kang, E.-S. An, H. K. Kim, M. J. Eom, M. Lee, C. Park, T.-H. Kim, H. C. Choi, B. I. Min, and J. S. Kim, ACS Nano **10**, 8888 (2016).
[10] L. Li, W. Wang, L. Gan, N. Zhou, X. Zhu, Q. Zhang, H. Li, M. Tian, T. Zhai, Adv. Funct. Mater **26** 45 (2016).
[11] S. De Palo, F. Rapisarda, and Gaetano Senatore, Phys. Rev. Lett. **88**, 206401 (2002).
[12] J. P. Eisenstein and A. H. MacDonald, Nature **432**, 691–694 (2004).
[13] J. J. Su and A. H. MacDonald, Nat. Phys. **4,** 799 (2008).
[14] D. Nandi, A. D. K. Finck, J. P. Eisenstein, L. N. Pfeiffer & K. W. West, Nature **488,** 481–484 (2012).
[15] Perali, D. Neilson, and A. R. Hamilton, Phys. Rev. Lett. **110**, 146803 (2013).

[16] D. Snoke, Science **298**, 1368 (2002).
[17] L. V. Butov, C. W. Lai, A. L. Ivanov, A. C. Gossard, and D. S. Chemla, Nature **417**, 47 (2002).
[18] L.V. Butov, A. C. Gossard, and D. S. Chemla, Nature (London) 418, 751 (2002).
[19] Wilson, Solid State Commun. **22**, 551 (1977).
[20] H. Cercellier, C. Monney, F. Clerc, C. Battaglia, L. Despont, M. G. Garnier, H. Beck, P. Aebi, L. Patthey, H. Berger, and L. Forró, Phys. Rev. Lett. **99**, 146403 (2007).
[21] S.A. Sunshine, J.A. Ibers Inorg. Chem. **24**, 3611 (1985).
[22] F. J. Di Salvo, C. H. Chen, and R. M. Fleming, J. Less-Common Met. **116**, 51 (1986).
[23] Y. F. Lu, H. Kono, T. I. Larkin, A. W. Rost, T. Takayama, A. V. Boris, B. Keimer and H. Takagi, Nat. Comm. **8**, 14408 (2017).
[24] K. Sugimoto, N. Maejima, A. Machida, T. Watanuki, and H. Sawa, IUCrJ **5**, 158 (2018).
[25] K. Seki, Y. Wakisaka, T. Kaneko, T. Toriyama, T. Konishi, T. Sudayama, N. L. Saini, M. Arita, H. Namatame, M. Taniguchi, N. Katayama, M. Nohara, H. Takagi, T. Mizokawa, and Y. Ohta, Phys. Rev. B **90**, 155116 (2014).
[26] Y. Wakisaka, T. Sudayama, K. Takubo, T. Mizokawa, M. Arita, H. Namatame, M. Taniguchi, N. Katayama, M. Nohara, and H. Takagi, Phys. Rev. Lett. **103,** 026402 (2009).
[27] K. Fukutami, R. Stania, J. Jung, E. F. Schwier, K. Shimada, C. I. Kwon, J. S. Kim, H. W. Yeom, Phys. Rev. Lett **123**, 206401 (2019).
[28] J. Lee, C.-J. Kang, M. J. Eom, J. S. Kim, B. I. Min, and H. W. Yeom, Phys. Rev. B **99**, 075408 (2019).
[29] K. Sugimoto, S. Nishimoto, T. Kaneko, and Y. Ohta, Phys. Rev. Lett. **120**, 247602 (2018).
[30] T. I. Larkin, A. N. Yaresko, D. Pröpper, K. A. Kikoin, Y. F. Lu, T. Takayama, Y. L. Mathis, A. W. Rost, H. Takagi, B. Keimer, and A. V. Boris, Phys. Rev. B **95**, 195144 (2017).
[31] T. I. Larkin, R. D. Dawson, M. Höppner, T. Takayama, M. Isobe, Y. L. Mathis, H. Takagi, B. Keimer, and A. V. Boris, Phys. Rev. B **98**, 125113 (2018).
[32] S. Mor, M. Herzog, J. Noack, N. Katayama, M. Nohara, H. Takagi, A. Trunschke, T. Mizokawa, C. Monney, and J. Stähler, Phys. Rev. B **97**, 115154 (2018).
[33] D. Werdehausen, T. Takayama, G. Albrecht, Y. Lu, H. Takagi, and S. Kaiser, J. Phys.: Condens. Matter **30**, 305602 (2018).
[34] S. Mor, M. Herzog, D. Golež, P. Werner, M. Eckstein, N. Katayama, M. Nohara, H. Takagi, T. Mizokawa, C. Monney, and J. Stähler, Phys. Rev. Lett. **119**, 086401 (2017).
[35] K. Okazaki, Y. Ogawa, T. Suzuki, T. Yamamoto, T. Someya, S. Michimae, M. Watanabe, Y.-F. Lu, M. Nohara, H. Takagi, N. Katayama, H. Sawa, M. Fujisawa, T. Kanai, N. Ishii, J. Itatani, T. Mizokawa, and S. Shin, Nat. Commun. **9**, 4322, (2018).
[36] T. Tang, H. Wang, S. Duan, Y. Yang, C. Huang, Y. Guo, D. Qian, and W. Zhang, Phys. Rev. B **101**, 235148 (2020).
[37] D. Werdehausen, T. Takayama, M. Höppner, G. Albrecht, A. W. Rost, Y. Lu, D. Manske, H. Takagi, and S. Kaiser, Sci. Adv. **4**, eaap8652 (2018).

[38] D. Werdehausen, S. Y. Agustsson, M. –J. Kim, P. Shabestari, E. Huang, A. Pokharel, T. Larkin, A. Boris, T. Takayama, Y. Lu, A.s W. Rost, H. Chu, A. Yaresko, M. Höppner, A. Schulz, D. Manske, B. Keimer, H.Takagi, S. Kaiser, Proc. SPIE 10638, Ultrafast Bandgap Photonics III; 1063803 (2018)
[39] P. Andrich, H. M. Bretscher, Y. Murakami, D. Golež, B. Remez, P. Telang, A. Singh, L. Harnagea, N. R. Cooper, A. J. Millis, P. Werner, A. K. Sood, A. Rao, arXiv: 2003.10799.
[40] B. Remez and N. R. Cooper, Phys. Rev. B **101**, 235129 (2020).
[41] Y. Murakami, D. Golež, T. Kaneko, A. Koga, A. J. Millis, and P. Werner, Phys. Rev. B **101**, 195118 (2020).
[42] Nakano, T. Hasegawa, S. Tamura, N. Katayama, S. Tsutsui, and H. Sawa, Phys. Rev. B **98**, 045139 (2018).
[43] T. Kaneko, T. Toriyama, T. Konishi, and Y. Ohta, Phys. Rev. B **87**, 035121 (2013).
[44] G. Mazza, M. Rösner, L. Windgätter, S. Latini, H. Hübener, A. J. Millis, A. Rubio, A. Georges, Phys. Rev. Lett. **124**, 197601 (2020).
[45] M. D. Watson, I. Marković, E. A. Morales, P. Le Fèvre, M. Merz, A. A. Haghighirad, and P. D. C. King, Phys.Rev.Res. **2**, 13236 (2020).
[46] A. Subedi. Phys. Rev. Materials **4**, 083601 (2020).
[47] J. Yan, R. Xiao, X. Luo, H. Lv, R. Zhang, Y. Sun, P. Tong, W. Lu, W. Song, X. Zhu, and Y. Sun, Inorg. Chem **58**, 9036 (2019).
[48] P. A. Volkov, M. Ye, H. Lohani, I. Feldman, A. Kanigel, K. Haule, G. Blumberg, arXiv:2007.07344.
[49] K. Kim, H. Kim, J. Kim, C. Kwon, J. S. Kim, B. J. Kim, arXiv:2007.08212.
[50] E. Baldini, A. Zong, D. Choi, C. Lee, M.H. Michael, L. Windgaetter, I.I. Mazin, S. Latini, D. Azoury, B. Lv, A. Kogar, Y. Wang, Y.F. Lu, T. Takayama, H. Takagi, A.J. Millis, A. Rubio, E. Demler, N. Gedik, arXiv:2007.02909.
[51] P. Andrich, H.M. Bretscher, P. Telang, A. Singh, L. Harnaga, A.K. Sood, A. Rao. arXiv:2007.03368.

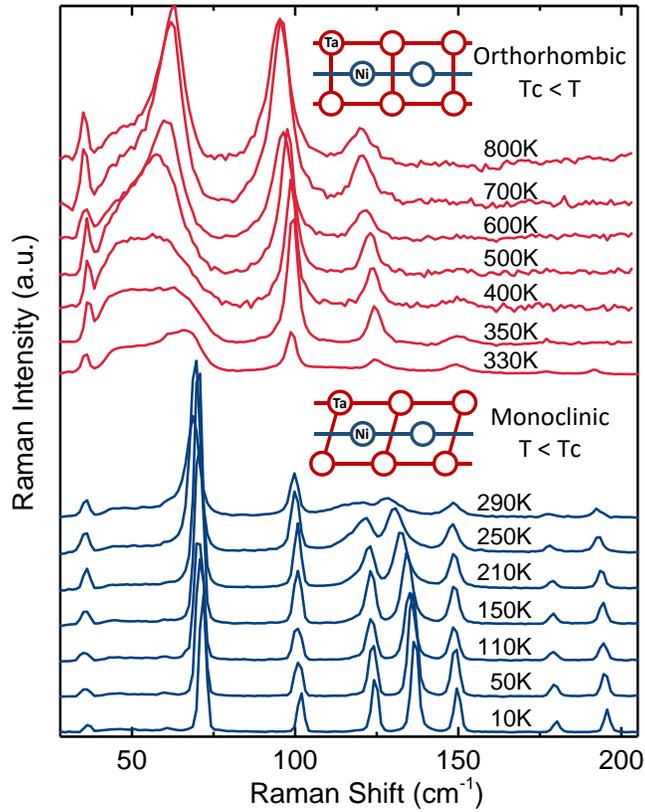

Figure 1. Raman spectrum of Ta2NiSe5 in the orthorombic phase above (red) and in the monoclinic phase below (blue) the structural phase transition at Tc=328 K. At the phase transition the Ta chains perform a shear displacement with respect to the Ni chain. The coressponding crystal structures are shown as insets. Spectra are taken in (X-) geometry revealing the Ag channel below and Ag+B2g channel above Tc.

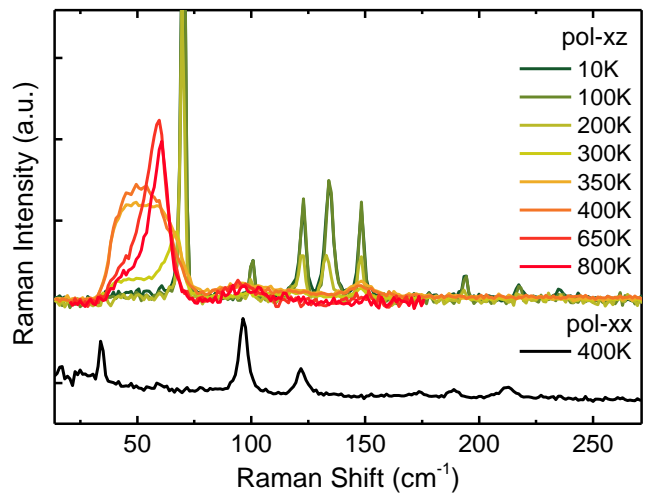

Figure 2. Temperature dependent Raman spectra in the (XZ) geometry probing the Ag channel below Tc and the B2g channel above Tc. The Ag channel above Tc is shown for 400 K in (XX) geometry.

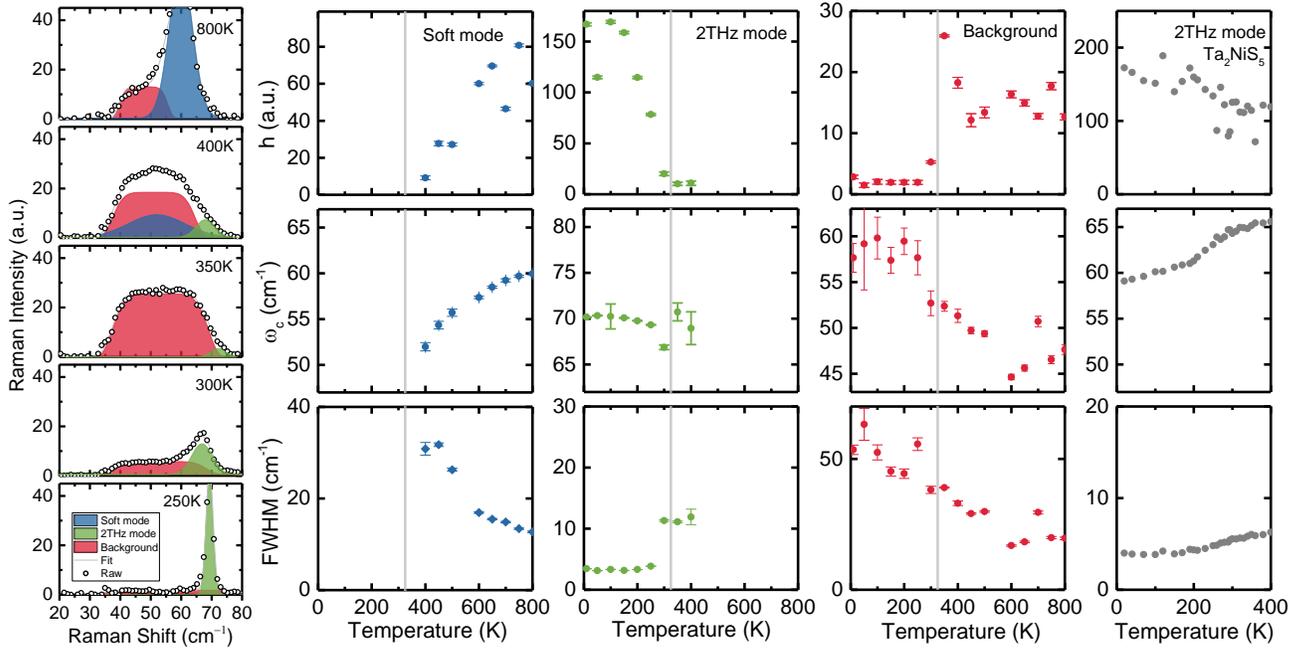

Figure 3. Fit to the low frequency Raman dynamics of Ta2NiSe5 in the (XZ) channel. Exemplary spectra and extracted parameters of amplitude center frequency and width of the different modes. The key contributions are (1) zone-center B2g soft phonon mode above Tc (blue), and (2) a 2 THz Ag mode (green) interpreted as amplitude modulation of the order parameter deriving from the unstable B2g mode. In addition (3) a broad spectral feature (red) next to these modes appears across the phase transition with its intensity becoming dominant at Tc. For comparison the dynamics of the 2THz mode in Ta2NiS5 are given which does not undergo any phase transition (grey).

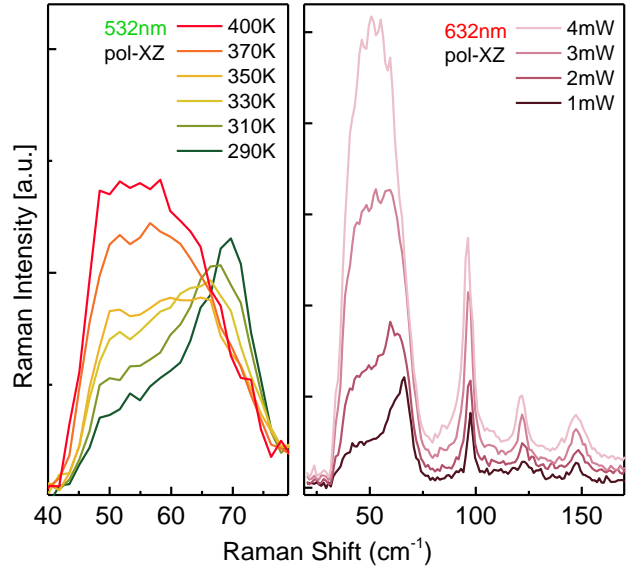

Figure 4. Laser dependent low frequency Raman dynamics of Ta2NiSe5. (a) Temperature dependent Raman spectra using 532nm excitation wavelength. (b) Laser power dependent Raman spectra at 300K using 632nm excitation. Spectra normalized to the increased fluence are shown in the supplemental material.

**Supplemental materials:**
**Observation of phononic soft mode behavior and a strong electronic background across the structural phase transition in the excitonic insulator Ta2NiSe5**

Min-Jae Kim[1,2], Armin Schulz[1], Tomohiro Takayama[1,3], Masahiko Isobe[1], Hidenori Takagi[1,3,4], and Stefan Kaiser[1,2]

[1] Max Planck Institute for Solid State Research, Stuttgart, Germany
[2] 4th Physics Institute, University of Stuttgart, Germany
[3] Institute for Functional Matter and Quantum Technologies, University of Stuttgart, Germany
[4] Department of Physics, University of Tokyo

mj.kim@fkf.mpg.de, s.kaiser@fkf.mpg.de

**S1: Methods**

Single crystals of Ta2NiSe5 were grown by chemical vapor transport reaction described in [23]. The linear Raman spectrum was measured using a Jobin Yvon Typ V 010 LabRAM single grating spectrometer, equipped with a double super razor edge filter and a Peltier-cooled charge-coupled device camera. The resolution of the spectrometer (grating, 1800 lines/mm) was 1cm−1. The spectra were taken in a quasi-backscattering geometry using the linearly polarized 632.817-nm line of a He/Ne gas laser. The power was lower than 1 mW, and the spot size was 20 μm. The scattered signal was filtered and analyzed using an additional polarizer before the spectrometer. The experimental configurations $Y(XX)\bar{Y}$ and $Y(XZ)\bar{Y}$ probe only Ag representations in the monoclinic phase (C2/c) and Ag, B2g channel respectively in the orthorhombic phase (cmcm) of TNSe. The $Z(YX)\bar{Z}$ configuration accesses the Bg channel in the monoclinic phase and B1g channel in the orthorhombic phase.

**S2: Raman response of Ta2NiSe5 in the different symmetry channels**

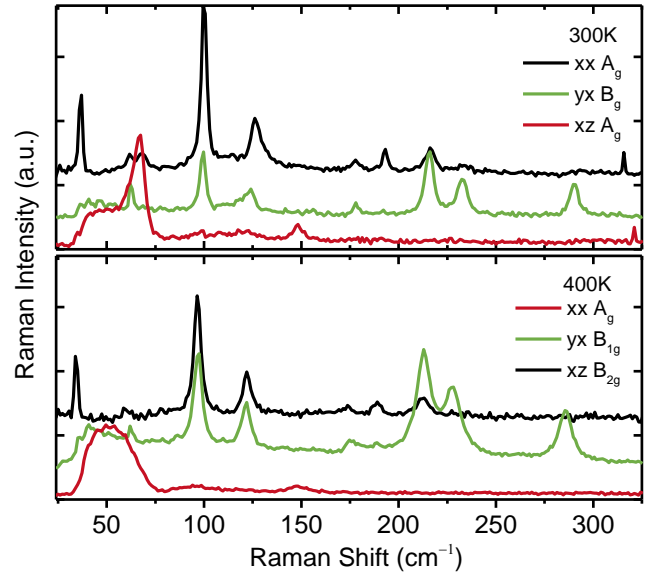

Fig. S2. Raman spectra of Ta2NiSe5 at 300 K in the monoclinic and at 400 K in the orthorhombic phase decomposed into different symmetry channels.

Figure S2 shows the polarization-dependent Raman spectra in the monoclinic phase below Tc at 300 K and in the orthorhombic phase above Tc at 400 K. In the XX configuration at 300K the well-known sharp Ag modes at 36, 101, 124cm-1 and the other high frequency modes are visible. The YX configuration of the Bg channel besides the Bg modes shows some leakage of peaks from the XX configuration because of the quasi quasi-one-dimensional crystal structure of the real system. But these are not in the focus of the present manuscript. Most important the XZ configuration does not show strong leakage of modes of other configurations. In this channel the 71 cm-1 Ag mode as well as the broad background feature are clearly visible. As mentioned in the main text the feature is present already for temperatures below Tc. Above Tc at 400K Ag and B1g channels in XX and YX geometries show the already known high frequency phonons. Notable is the XZ channel that probes the B2g channel above Tc. In the low frequency range the sharp peak at 71cm-1 disappears and the remarkable broad background feature arises above Tc.

**S3: Laser heating**

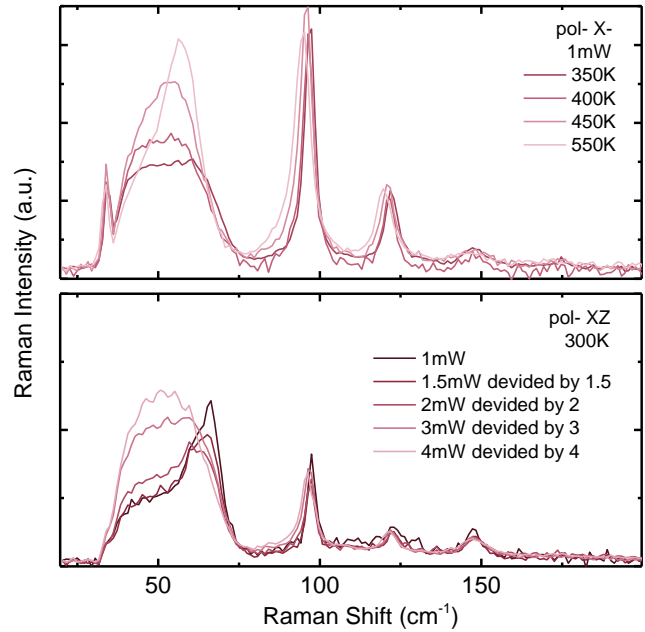

Fig. S3. (a) Thermal heating of Ta2NiSe5 at 1 mW laser power. (b) Laser heating with normalized power dependent Raman spectra for 1, 1.5, 2, 3, and 4 mW laser power at 300 K.

Figure S3 compares the different response of Ta2NiSe5 to thermal or laser heating. Figure S3 (a) shows the thermal heating where the laser power stays fixed. Most prominent heating effects are the dominant change of the broad background feature and the soft mode behavior of the B2g mode as described in the main text. In stark contrast, Figure S3 (b) shows the power dependent Raman spectra at fixed base temperature of 300K. The data is normalized to the photon flux of the base fluence at 1mW. We compare the (X-) with the (XZ) configuration since in the laser heating data the modes at 96 and 121 cm-1 are Ag modes leaking from the (XX) channel. As seen at the phonon modes at 96, 121 and 150cm-1 there is only a minor heating effect on the lattice. Only a small increase and broadening can be seen on the 96 cm-1 mode and only a small frequency shift compared to thermal heating can be detected. On the contrary the broad background feature around 50 cm-1 and the 71 cm-1 Ag mode show a heating effect that is highly susceptible to the increasing laser fluence. We find a prominent increase the broad feature that becomes dominant compared to the phonon modes. This behavior we attribute to electronic heating by the laser fluence and therefore attribute an electronic or strongly joint electronic-phononic origin to the broad background feature. Its response peaks around Tc as described in the main text. That is in line with the interpretation that the feature is related to excitonic fluctuations as discussed in other recent Raman studies [48,49] where a low frequency electronic background of excitonic fluctuations is also found to peak at Tc. [49]. Interesting to note is that the 71 cm-1 Ag mode seems to be fully suppressed by the electronic feature. However no striking onset of the soft mode B2g phonon is found that seems to be fully buried under the electronic feature or also is suppressed when the system heats up with higher laser fluences. The strong influence of the electronic heating of the fluctuation feature to the 71 cm-1 Ag mode below Tc and the B2g soft mode in the high temperature range show the peculiar coupled response of structural and electronic order that drives the phase transition and that we characterized in Figure 3 of the main text. The striking difference between electronic and lattice heating call for more detailed studies in future. Maybe higher laser fluences are needed to heat the full lattice system and make a potential soft B2g soft mode also visible. Comparing to the data with very high local laser fluence and high temperatures in Ref [49] that reports no soft mode behavior strongly suggests that the soft mode behavior is suppressed on laser heating. The different response of the high frequency Ag modes from the 71 cm-1 Ag and the B2g mode also shows that the latter two modes are coupled to the excitonic fluctuations compatible with a picture of self localized excitons in polaronic complexes as proposed in [30,31].

**S4: Raman spectra of Ta2NiSe5 wit Bose correction factor and different experimental geometries**

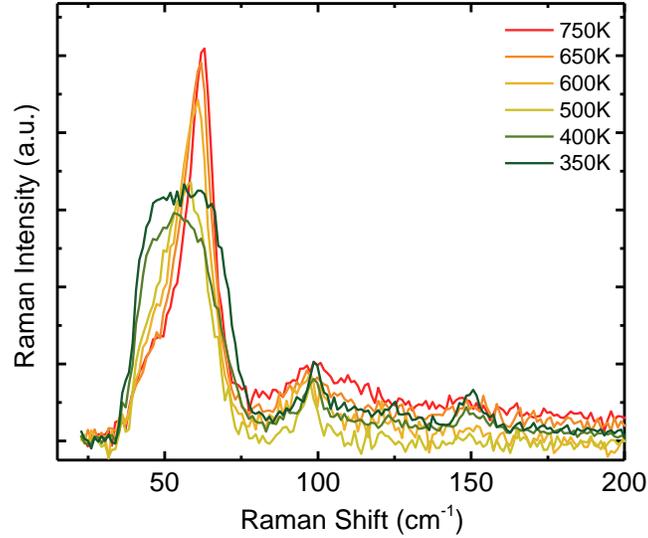

S4.1 Bose corrected Raman spectra of (XZ) geometry in Ta2NiSe5.

To compare our data to other Raman studies that recently appeared as preprints [48,49] we also show our data Bose factor corrected as done in these studies. As seen in Fig. S4.1 the features discussed in the main text remain independent of the correction factor. In particular the broad electronic background in the B2g channel in the orthorhombic phase and the B2g soft mode behavior remain distinct and dominant features. No broadband continuum arises that would require us to fit a strong Fano coupled response of the phonons with the electronic background. A key difference between the studies [48,49] are the applied laser power for the Raman studies and possible leakage from other polarization directions. In particular the study [49] uses a local power per area that is more than 100 times larger than in our present study. The study [48] uses a similar fluence as we do in our study for the low temperature regime but uses laser heating to reach the high temperatures across the phase transition. As discussed in Fig. 4 and S3 laser heating has a strong influence on the dynamics of the system since the electronic and phononic system are not heating up equally.

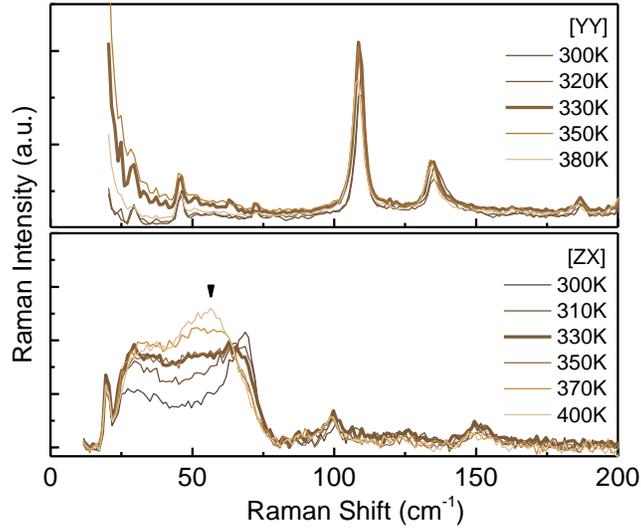

S4.2 Low frequency Raman dynamics in the (a) (YY) channel and (b) (ZX) channel in Ta2NiSe5.

A divergent low frequency background like in Ref. [49] we find only for a different sample geometry (figure S4.2): (YY). We suspect that part of the background in [49] may come from leaking of other crystallographic directions. In addition strong electronic heating effects apply for the high fluence used in this study. As discussed this dominates the response and covers the soft phonon response in the study [49]. However, the electronic background they find peaks around Tc, similar to our electronic dominated background feature discussed in Figure 3 showing the strong impact of excitonic fluctuations across the phase transition.

We can find a similar background as in Ref. [48] with a hump around 30 cm-1 if we probe in (ZX) geometry (figure S4.2). In contrast to Ref. [48] the clear B2g soft mode remain. Ref. [48] does not see this soft mode behavior because they heat the sample with high laser power rather than using a thermal heater. As we have discussed and shown in Figure 4 and S3 this heats dominantly the electronic system and suppresses the appearance of the B2g soft mode. Therefore they only observe the excitonic response that is highly susceptibly to the laser fluence.

Comparing the present Raman studies we conclude that indeed a strong excitonic response is present in the system and participates at the phase transition. In our study we clearly see this in form of the strong electronic background feature. However, as we have shown this electronic background becomes dominant for high laser fluences. Already relatively low power of the excitation laser is sufficient to suppress the soft mode behavior of the phononic system that we report here and therefore is absent in both of the other studies. In addition our study reaches a much higher temperature regime (even without substantial laser heating) than both of the other Raman studies where the B2g soft mode behavior becomes a clear and distinct feature at temperatures above 400 K.